\documentclass[12pt]{iopart}
\usepackage{hyperref}
\usepackage{listings}
\usepackage{graphicx}
\expandafter\let\csname equation*\endcsname\relax
\expandafter\let\csname endequation*\endcsname\relax
\usepackage{amsmath}
\usepackage[sorting=none]{biblatex}
\bibliography{papers}
\usepackage{enumitem}
\usepackage[separate-uncertainty=true]{siunitx}
\usepackage{verbatim}

\usepackage{tcolorbox}
\usepackage{newfloat}
\DeclareFloatingEnvironment[fileext=lop]{example}

\newcommand{\dv}{davinci-003 }
\newcommand{\cg}{ChatGPT }
\newcommand{\sample}[4]{
\begin{example}[htp!]
\begin{tcolorbox}
\textbf{\flushleft Prompt: #1}
\\ \\ Output:#2\\%}
\end{tcolorbox}
\caption{#3}
\label{#4}
\end{example}
}
\renewcommand{\theenumi}{\arabic{enumi}}
\DeclareUnicodeCharacter{03B1}{$\alpha$}
\DeclareUnicodeCharacter{03B2}{$\beta$}
\DeclareUnicodeCharacter{03B3}{$\gamma$}

\begin{document}
\title{The Death of the Short-Form Physics Essay in the Coming AI Revolution}
%\author{anonymous}
\author{Will Yeadon, Oto-Obong Inyang, Arin Mizouri, Alex Peach, Craig Testrow}
%\address{anonymous}
\address{Department of Physics, Durham University, Lower Mountjoy, South Rd, Durham, DH1 3LE, UK}
%\ead{anonymous}
\ead{will.yeadon@durham.ac.uk}
\vspace{10pt}
\begin{indented}
\item[]{December 2022}
\end{indented}

\begin{abstract}
The latest AI language modules can produce original, high quality full short-form ($300$-word) Physics essays within seconds. These technologies such as \cg and \dv are freely available to anyone with an internet connection. In this work, we present evidence of AI generated short-form essays achieving first-class grades on an essay writing assessment from an accredited, current university Physics module. The assessment requires students answer five open-ended questions with a short, $300$-word essay each. Fifty AI answers were generated to create ten submissions that were independently marked by five separate markers. The AI generated submissions achieved an average mark of $71 \pm 2 \%$, in strong agreement with the current module average of $71 \pm 5 \%$. A typical AI submission would therefore most-likely be awarded a First Class, the highest classification available at UK universities. Plagiarism detection software returned a plagiarism score between $2 \pm 1$\% (Grammarly) and $7 \pm 2$\% (TurnitIn). We argue that these results indicate that current AI MLPs represent a significant threat to the fidelity of short-form essays as an assessment method in Physics courses. 
%We present a double blinded study comparing the marks received for short-form (300 word) essays written by second year students at Durham University and by a free to use AI model - the GTP-3 model \dv. We demonstrate how AI ..... the performance of students. Critically, we found the AI written essays pass modern plagiarism detection software at a higher/lower rate than student work making them extremely difficult to detect. The results suggest that short-form essays are no longer suitable as an assessment mechanism.
\end{abstract}

\begin{section}{Introduction}\label{intro}
\subsection{Background}
\label{subsection-bkgd}
% This paper is for \PED
AI text-completion technologies have undergone rapid development in recent years, outpacing commonplace perceptions of their quality. It is now possible to use these tools to reliably produce quality content that is accurate, clear and critical, on practically any topic of choice. These tools can produce content that is highly idiosyncratic, and which is nearly indistinguishable from human-produced content \cite{gpt-3-indistinguish, NYT-gpt}. Furthermore, these technologies are becoming increasingly accessible, fast, cheap and incredibly easy to use, essentially only requiring users to specify a brief text prompt. The effort required to produce such a prompt could entail simply copying and pasting a question from an assignment. Clearly, such technologies could pose a threat to the fidelity of existing forms of assessment, as AI written work could be submitted by students and both pass undetected through plagiarism prevention software and score higher marks than the student could have had they written the work themselves. In this work, we provide evidence that this threat is very significant. We demonstrate how high quality short-form essays, written by AI software in only a few seconds, can score a First Class for an assignment from an accredited university Physics module. This, we argue, effectively renders the short-form essay obsolete as an assessment tool. Despite popular reservations, the AI revolution has begun, and it is vital that educators remain proactive and aware of the capabilities of these technologies and their potential ramifications for teaching and assessment, particularly at this very early stage in the development of AI text-completion technologies. 

Modern systems such GPT-3 are already extremely proficient at producing accurate, insightful responses on almost any topic, with applications in diverse areas such as healthcare \cite{SA-medical-qa}, legal advice chat bots \cite{legalAdvice} and providing feedback for student's work \cite{student-feedback}. They are thus potentially capable of answering the short-form essay questions which form the basis of assessments in many university modules. Whilst there are some technologies capable of detecting AI generated text \cite{detect-ai}, universities currently only look for plagiarism rather than whether text was AI generated. Previous technological innovations such as the widespread use of virtual learning within universities have had muted impact on teaching practices \cite{VLEimpact} yet the current capabilities of relatively earlier stage technologies such as \cg and \dv is likely to change this for the case of AI.

Given technologies such as \dv and \cg are only months old, there is limited prior work on AI written essays. An older language model, GTP-2, has been used by students to write essays in a co-piloting role \cite{aiCoPilotEssays}. Here, Students were tasked with blending an essay they have written themselves with output from GPT-2. However, new models such as \cg and \dv can write entire essays using a single, user-defined sentence prompt. In fact, through stringing together several prompts GTP-3 has already demonstrated the potential to create an entire academic paper \cite{gpt3AcademicPaper}. If one only cherry-picks the most exemplary outputs as examples of AI capabilities, this invariably implicates survivorship bias. However, an hypothetical student who wished to use an AI tool to answer their essay would most-likely choose the best output they could from a set of multiple, rapidly-generated options. Clearly it is prudent to appraise the capabilities of these technologies within the context of how they will most likely be used by students.

\subsection{GPT-3: \dv and \cg}
\label{subsection-gpt3}
OpenAI's GTP-3 model is an autoregressive language model \cite{gtp3-paper} that processes and generates text. Trained on a large dataset of human-generated text, GPT-3 can analyze a prompt (the input text) and use statistical techniques to predict the most likely next word or words that form a coherent and meaningful statement. Within the GTP-3 group of NLP models, \dv was released in late November 2022. This was followed by \cg in December 2022. OpenAI's website features a free 'playground' web application which open allows a user to enter a short prompt from which \dv will generate additional text. The 'playground' web application features various parameters a user can adjust such as the maximum length of the output or how random it is. Conversely, \cg is a chatbot that seeks to converse with the user via text. Both technologies can produce text output but \dv was principally used in this study. 

As an example of the sophistication of the \dv output, Example \ref{expLit} shows the output for a question demanding literary analysis of William Shakespeare. Note here how, whilst simple, this prompt still demands recognizing Shakespeare as a name of a person whilst Summer and Winter are recognized as names of seasons. Here, Summer is associated more closely to happiness and Winter is associated closer to darkness. Darkness is not an antonym of happiness.

\sample{Explain why Shakespeare compared someone to a Summers’ day instead of a Winters’ day.}{ Shakespeare most likely compared someone to a summers' day instead of a winters' day because a summers' day is typically seen as a day of happiness, while a winters' day is seen as a day of darkness.}{\dv output showing relatively straightforward literary analysis.}{expLit}

Whilst these kind of NLP models have been around for some time, \dv provides output which demonstrates the critical understanding and reason required to create excellent answers to essay questions. In Example \ref{expMoral}, \dv appears to be considering the moral implications of using AI technologies to generate essays. This is despite how the text is generated based on learned patterns and structures.

\sample{Is it ethical to use AI text-completion technologies for your homework assignments?}{ There is no right answer to this question since it can be seen as ethical or unethical depending on one's perspective. Some people might see it as unethical because it is cheating to use AI technologies to complete homework assignments. Others might see it as ethical because it is simply using a tool to help complete the assignment. Ultimately, the decision of whether or not to use AI technologies for homework assignments is up to the individual.}{\dv output that appears to take a moral position.}{expMoral}
 
\end{section}
\begin{section}{Method}
\subsection{Outline of the module and exam}
Physics in Society is a module offered by the Department of Physics at Durham University. The module content consists of the history and philosophy of Physics, the development of modern science and the ethical issues surrounding the use of technology in society. The major assessed component of the module is the exam which consists of a set of five short-form essay questions, each of no more than $300$ words, on a selection of topics from the module. Figure \ref{fig-form-qs} shows the five questions used for the AI submission generation. These questions serve to test a variety of aspects such as the history, philosophy, communication and ethics of Physics. This is reflected in the module assessment proforma, shown in Figure \ref{fig-rubric}, that outlines the five key criteria against which essays are marked.

\begin{figure}[h]
    \centering
    \begin{tcolorbox}[
    colframe=black!25,
    colback=yellow!10]
    \begin{enumerate}
          \item Is Physics based on facts that follow from observations?
          \item What was the most important advance in natural philosophy between 1100 and 1400?
          \item How did natural philosophers' understanding of electricity change during the 18th and 19th centuries?
          \item Does Kuhn or Popper give a more accurate description of physics?
          \item Was there a scientific revolution in 17th-century Europe?
    \end{enumerate}
    \end{tcolorbox}
    \caption{The five questions used to generate submissions. Note a variety of historical, philosophical and factual themes are covered over the $5$ questions.}
    \label{fig-form-qs}
\end{figure}

\begin{figure}[h]
    \centering
    \begin{tcolorbox}[
    colframe=black!25,
    colback=yellow!10]
    \begin{enumerate}
        \item{Is there a high academic content, at a suitably advanced level, thus indicating that the student is familiar with some of the key milestones in the history of physics, the philosophy of physics, science communication or ethics in academia?}
        \item{Has the student formed an appreciation of the physics underlying a particular topic? }
        \item{Does the student demonstrate a thorough grasp of the subject material? }
        \item{How well does the student address the specific question asked? }
        \item{Is the work written in a suitably authoritative, academic style, with the material presented in a logical, coherent and concise manner and supported by appropriate factual information?}
    \end{enumerate}
    \end{tcolorbox}
    \caption{The Physics in Society assessment proforma used to grade submissions.}
    \label{fig-rubric}
\end{figure}

Students are awarded a mark between \emph{0} and \emph{100} for each of the five categories specified in the module assessment proforma which are shown in Figure \ref{fig-rubric}; each of these categories is equally weighted when determining the students' final mark. The mark awarded for each category is based on the answers for all five questions; hence the marks must be balanced and reflect a general level of consistency within the answers. Students typically score $71 \pm 5$ on the Physics in Society module \cite{hall_2021}.

\subsection{Generating the AI scipts}
\label{subsection-sample-gen}
A sample of $n=10$ AI generated scripts, each containing five question-answer pairs, was compiled from a set of outputs of \dv, generated with suitably-chosen prompts based on the questions given in Figure \ref{fig-form-qs} entered into the OpenAI 'playground' web application. Repeatedly inputting prompts consisting of direct quotes the questions can lead to GPT-3 responding briefly and laconically. More discursive, original responses can be generated by slightly rephrasing prompts. For example the prompt, \emph{'Was there a scientific revolution in 17th-century Europe?'} could be re-phrased as\emph{'Did 17th-century Europe experience a scientific revolution?'}, or \emph{'Was there a scientific revolution in seventeenth-century Europe?'}, which will generate more original responses. Rephrasing can also be combined with prefixing the prompt with phrases such as \emph{'Write a long and detailed essay on'}, \emph{'In 300 words or more, comment on'} or \emph{'Explain in detail using more than 250 words'}. Such prompts are effective at instructing the AI to consistently generate more discursive responses. Here, the stated number of words used can be changed in tandem with the maximum output setting on the 'playground' to induce further novelty. Additionally, one can easily instruct GPT-3 to argue for (against) a certain proposition, by choosing a suitable prompts such as, \emph{'Explain why there was (not) a scientific revolution in 17th-century Europe'}. This can be exploited effectively in cases where there are many viable answers to pick, but GPT-3 often favours a particular one. For example, the prompt \emph{'What was the most important advance in natural philosophy between 1100 and 1400?'} typically generates a response from GPT-3 indicating that the scientific method was the most important advance. However, if the user explicitly states a relevant historical figure - such as Aquinas, Bacon or Buridan - within the prompt, as with the following example prompt; \emph{'Explain why Thomas Aquinas' work was the most ...'} then GPT-3 will now provide a response that is consistent with the proposition specified by the user.

To obtain a consistent, minimal and fair benchmark of the quality of the \dv essays, the output was not edited in any way. Only if the output was excessively similar to a prior generated one would it be rejected entirely, and a new one generated. It is reasonable to assume that a student who wished to use \dv to write their answers would apply some minor edits, yet even without this the AI answers were of excellent quality. An example of \dv output is presented in Example \ref{outputRaw}.

\sample{Write a 300 word essay explaining why physics is not based on facts that follow from observations.}{ Physics is not based purely on facts that follow from observations because some aspects of physics go beyond the reach of observation. At the root of this is the fact that physics is a field of study that attempts to uncover the fundamental workings of the universe. Physics strives to answer questions about how things work on the most basic level--questions about the behavior of particles and forces, for example. It is impossible to directly observe most of these phenomena; when an experiment is conducted, the results are generally observed indirectly through the measurements of related quantities. Therefore, some aspects of physics are based on theoretical models that do not rely directly on observable facts, but instead on the assumptions contained in the models.
\\\\
Another reason why physics is not based on facts that follow from observations is that observations are always open to interpretation based on the observer’s perspective. In general, different observers will have different interpretations of the same observations. This can lead to disagreement and debate over the interpretation of observed phenomena, which is not conducive to building a consensus on the truth of physical laws or theories.
\\\\
Finally, even when multiple observers agree on the interpretation of an observation, there is no guarantee that the conclusion is correct or that it can be generalized to other physical phenomena. In other words, observations are often limited in terms of the conclusions that can be drawn from them. This means that while they are an important part of physics, they do not form the sole basis of the field.
\\\\
In conclusion, physics is not based solely on facts that follow from observations because some aspects of physics go beyond the reach of direct observation. Additionally, observations are open to interpretation and are often limited in terms of the conclusions that can be drawn from them. Thus, while observations are an important part of the physics, they do not form the sole basis for the field.}{Unedited 328-word output from \dv using a temperature (randomness) of $0.95$ and a maximum output length of $700$ tokens where $1$ token is roughly $4$ characters of English text.}{outputRaw}
\end{section}

\begin{section}{Analysis and results}
\label{section-analysis}
\subsection{Overview}
\label{subsec-analysis-overview}
The ten AI generated submissions were marked independently by five different markers and achieved an average mark of $71 \pm 2 \%$. This compares to the average of $71 \pm 5 \%$ in Physics in Society and an average score of $72 \pm 3 \% $ for all second year Physics modules at Durham University \cite{hall_2021}. These results strongly imply that \dv and \cg can write short-form Physics essays at a level of competency that is comparable to second year Physics students. 

A significant trait of the AI generated answers were their consistent quality, as seen in the tight distribution of AI scores compared with student scores for the module. Figure \ref{fig-ai-scores} reveals that the very top performing students are able to outscore the AI essays, but the majority of the cohort performs either comparably or worse. Our results imply that students in the bottom third of their cohort would have a significant incentive to submit AI generated work for their assignments instead of their own original work. 

Plagiarism detection tools cannot be relied upon to indicate the presence of AI generated work. This is shown in Table \ref{plagPer}, the AI essays score $2 \pm 1\%$ plagiarism on Grammarly and $7 \pm 2\%$ on TurnitIn. Both of these scores would be deemed sufficiently original if submitted in an actual university assessment; the only text highlighted as non-original were the questions given in Figure \ref{fig-form-qs}. Figure \ref{fig-marker-scores} depicts how consistent the independent markers were with their scoring, whereby they awarded $73.0 \pm 1.6$, $72.6 \pm 2.0$, $69 \pm 2$, $70 \pm 2$ and $70.6 \pm 1.9$. These scores are in strong agreement with the module average of $71 \pm 5$ and - combined with passing the plagiarism check - suggest it would be challenging to identify the AI essays from those of students. Furthermore, the average scores awarded by each marker are all in reasonable agreement with one another, indicating that the marking process is valid.
% Teasing next paper
\begin{table}
\centering
\caption{\label{plagPer} Table showing plagiarism percentage identified through both Grammarly and TurnitIn for each of the ten submissions.}
\begin{tabular}{@{}llllllllllll}
\br
Submission & \#1 & \#2 & \#3 & \#4 & \#5 & \#6 & \#7 & \#8 & \#9 & \#10 & Average \\
\mr
TurnitIn (\%) & 9 & 8 & 7 & 7 & 6 & 6 & 7 & 9 & 5 & 10 & $7 \pm 2$\% \\
Grammarly (\%) & 1 & 2 & 1 & 0 & 2 & 3 & 1 & 2 & 1 & 2 & $2 \pm 1$\% \\
\br
\end{tabular}
\end{table}

\begin{figure}[h]
    \centering
    \includegraphics[width = 13.5cm]{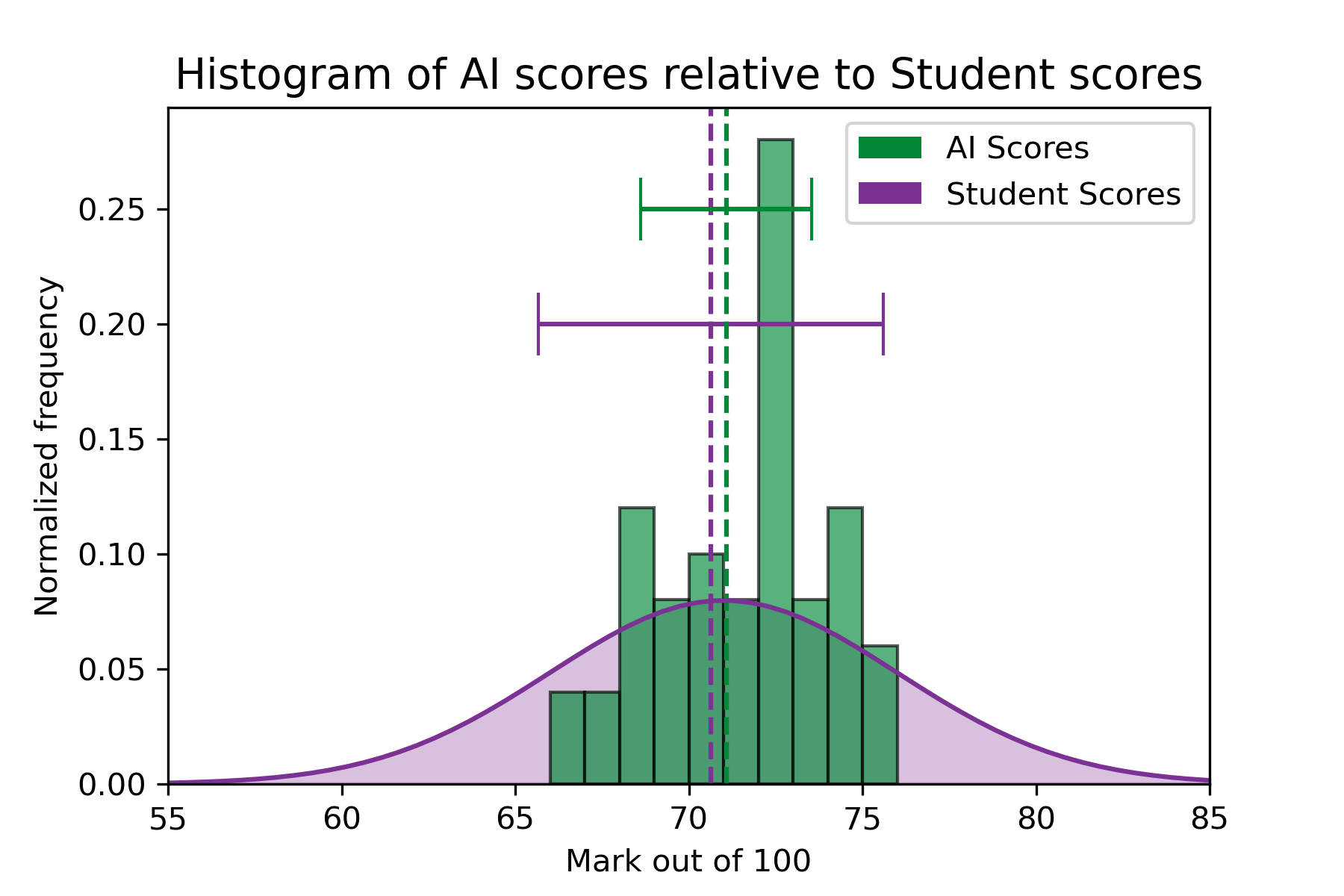}
    \caption{Histogram of AI scores (green) against a Gaussian distribution reflecting the average scores of students in Physics in Society \cite{hall_2021}.}
    \label{fig-ai-scores}
%\end{figure}
%\begin{figure}[!htb]
    \includegraphics[width = 13.5cm]{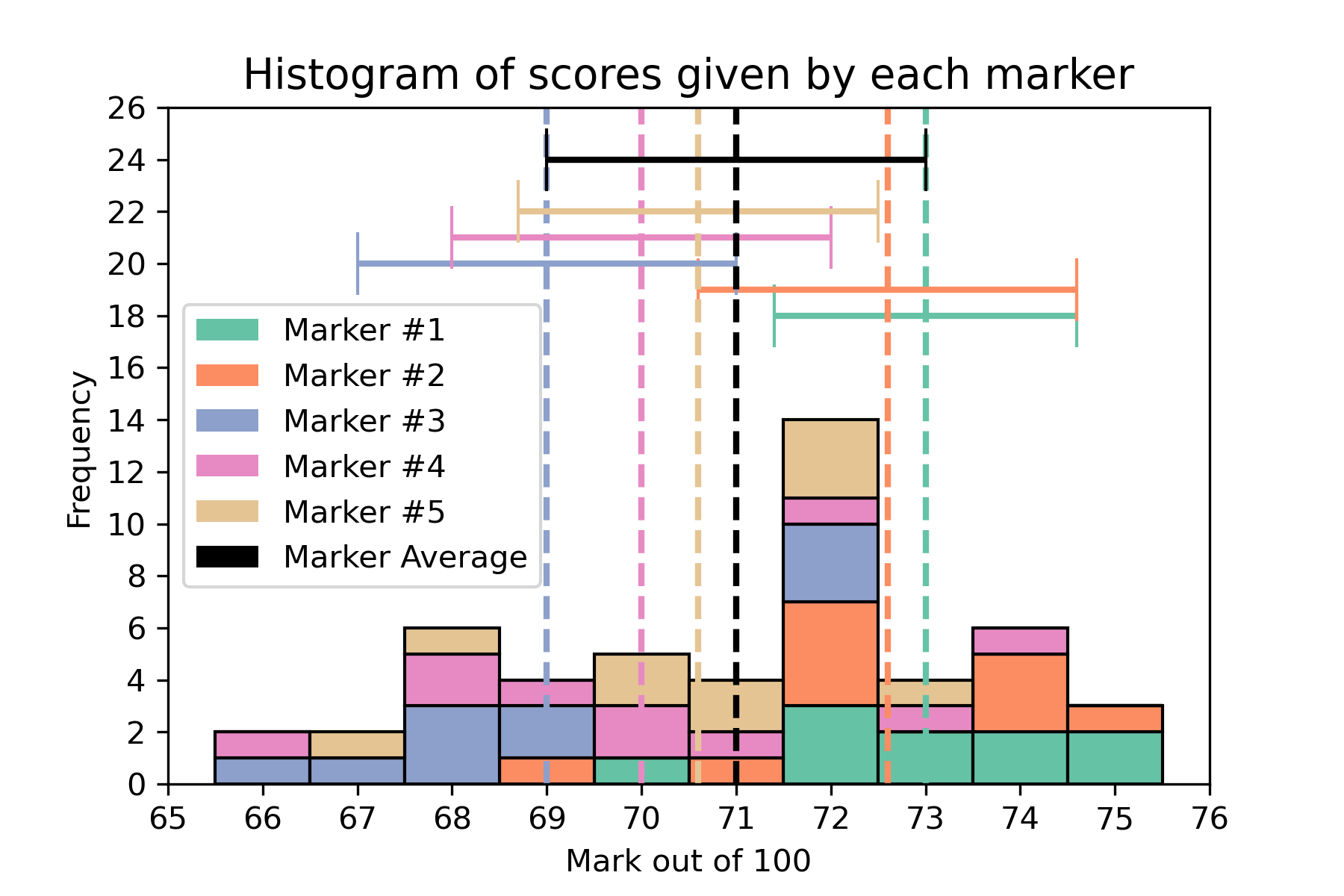}
    \caption{Stacked histogram of the independent marker's scores for the ten AI generated essays. The average and standard deviation for each marker is shown by the dashed line and error bars respectively. The overall average of the independent markers of $71 \pm 2$ is shown in black. Note the overlap between the different independent marker's scores.}
    \label{fig-marker-scores}
\end{figure}

\subsection{Written quality}
Looking closely at Example \ref{outputRaw} reveals how the readability of certain phrases could be improved. As an illustration the sentence '\emph{Physics strives to answer questions about how things work on the most basic level questions about the behavior of particles and forces, for example. It is impossible...}' could be rearranged to not finish with '\emph{for example}'. Similarly, '\emph{level--questions}' should be changed to '\emph{level questions}. In other essays often American English spelling was used - this would also require change for a UK context. These slight flaws show that the raw output is not necessarily semantically perfect. However, as prior mentioned, it is reasonable to assume that a student would apply at least some editing to the text if they wanted to use \dv to write their submissions for them. Thus this semantic imperfection doesn't stop the use of \dv as an essay writing tool. 
\end{section}

\begin{section}{Discussion \& Conclusion}
\label{section-disc}

\subsection{Impact on Higher Education}

The results of this short study suggest modern AI systems such as \dv and \cg are capable of writing high quality Physics essays, capable of achieving a first-class score, for free and within a few seconds. These tools essentially represent a wide spread democratization of paid essay writing services whereby anyone can have \dv or \cg write an essay for them. We could argue that, in contrast to merely sounding a death-knell for certain kinds of assessments, this technology will once again force us to re-think assessment, and will confer far greater benefits in the end than the comparatively small drawback of having to redesign existing assessments. Obviating the abuse of these technologies to undermine assessments could be as simple as requiring the students to produce the work within an invigilated setting. In such cases, the moral of the story is that while the threat to authentic assessment is profound, simple and practical solutions likely already exist that won't necessarily require a total upheaval of current assessments themselves, but may force a change of context at least.

Besides assessment, this technology allows users to generate innumerable, original examples, which can be shown by teachers within writing workshops. Similarly, as technologies such as \cg mature we could see their use as one-on-one tutors. Although, the availability of knowledge is effectively already widely democratized. The full course contents for most degrees at Stanford and MIT have been available on their respective MOOC platforms for a decade. Yet enrolling as a student at both institutions is still very competitive and a typical online MOOC can expect a completion rate of 10\% \cite{MOOCcompletion} which is far below that of a typical university degree. Thus it is not often clear in advance how new technology will impact Higher Education. 

\subsection{Ease of sample generation}
Getting \dv or \cg to respond in the desired manner often involves a process of trial-and-error, retrying or rewording prompts. Yet a transformational aspect of these technologies is how few times this needs to be done and how quickly the samples are generated. As covered in section \ref{subsection-sample-gen} it is relatively straightforward to get a good quality answer to an essay question simply by preceding the question with \emph{'Write a 300 word essay on'}. However, whilst $50$ AI answers were created, it is unclear whether additional unique responses could realistically be created without some familiarity with the subject. Take for example the question \emph{'What was the most important advance in natural philosophy between 1100 and 1400?'}, as covered in section \ref{subsection-sample-gen} it was possible to create multiple unique answers through rephrasing the question with a sensible suggestion. Rephrasing the question as  \emph{'Explain why X was the most important advance in natural philosophy between 1100 and 1400.'} where $X$ is Bacon’s work on optics / Buridan’s work on impetus / Wallingford’s astronomical clock or similiar allows for many unique answers to be created. However, this is contingent on knowing who Bacon / Buridan / Wallingford are in the first place. Yet simply knowing Roger Bacon worked on optics is not equivalent to being able to write a high quality essay on his work. Even if the phase-space of AI answers for questions could be checked by assessors beforehand, through partially answering the question and incorporating AI as a co-pilot students could still have AI write their answers but with a reduced risk of being caught. Secondly, as with paid essay writing services, without clear conclusive proof it is unreasonable to accuse a submission of not have being written by its stated author.  

\subsection{Limitations and Future Work}
In the present work, we have adopted a fairly rudimentary proxy, consisting of the average scores for the module, in order to contextualise the AI scores. More properly, we should compare the AI scores to those of the students for this exam. We aim to address this in a forthcoming paper in which we compare the performance of AI to student submissions, during which we also address the question of the extent to which markers can reliably distinguish between AI submissions and human submissions. Nevertheless, our key result, that AI can generate content that attains highly, in absolute terms, for short-form essays, is very significant in and of itself, despite the aforementioned limitations.

The focus of the present work are assignments consisting of short-form essay questions, which represent a minority of assessments within STEM teaching. However, future work will consider the capabilities of AI MLPs with regard to scientific report writing and answering analytical questions which require the use of calculations, coding, symbolic manipulation and algebraic typesetting, which are staples of STEM assignments and examinations. Remarkably, GPT-3 is already demonstrating a capacity for providing accurate, and properly formatted answers to these types of assignments, though this is beyond the scope of the present work. In our view, it is only a matter of time before these technologies develop to the point where they can not only pass, but excel in all of these aspects of STEM assignments too. 

\subsection{AI generated feedback for AI essay.}
\label{AI-feedback-ai}
Given the ability of AI to create high quality essays the next logical step is to get an AI to mark the AI essays. \cg is better suited to this task than \dv and the output of a \cg conversation is shown in Example \ref{expFeedback}. Whilst when scoring itself the AI is a little generous - awarding $85/100$ - it does provide specific feedback comments that directly relate to both the input essay and the rubric. 

\sample{Using the following rubric: (\emph{inserted rubric shown in Figure \ref{fig-rubric}}). What score, out of 100, would you give the following essay (\emph{inserted essay shown in Example \ref{outputRaw}}).}{ Based on the provided rubric, I would give the essay a score of 85 out of 100.
\\\\
The essay demonstrates a high level of academic content and an appreciation of the physics underlying the topic. The student thoroughly grasps the subject matter and addresses the specific question asked in a clear and concise manner. The essay is written in an academic style and presents the material in a logical, coherent manner, supported by appropriate factual information.
\\\\
One potential area for improvement would be to further discuss the limitations of observations and their impact on the development of physical theories. The essay touches on this, but it could be further developed to provide a more in-depth analysis of the role of observations in physics. Additionally, citing sources or referencing relevant literature could also strengthen the essay and add credibility to the arguments presented. Overall, the essay is well-written and provides a thorough analysis of the topic.}{\cg generated feedback for the AI essay in Example \ref{outputRaw}.}{expFeedback}

This concept of AI marking is a non-trivial point. Automatic essay scoring (AES) is a mature technology in regular use for US High School essays \cite{essayAutoMarkSherman, autoEssayLitReview}. These AES technologies are trained on many examples of student essays and assign a grade based on various factors such as grammar, spelling, word choice, organization, and content. This is in contrast to a university module like Physics in Society that features questions relatively unique to the content taught and where a cohort typically ranges from $50$ - $70$ students so there are limited examples for comparison. Despite this, the score and feedback given in Example \ref{expFeedback} is clearly cogent and would likely pass undetected if given to the student as human-written feedback. 

\subsection{Concluding thoughts}
We may be at the beginning of an AI revolution. In order to facilitate authentic assessment it is vital that we are aware of the capabilities of this technology and its ramifications on the way that credited work is assessed. In the present case, it is hard to avoid the conclusion that non-invigilated assessments based on short-form essay questions are already no longer fit for purpose; they are simply too vulnerable to current AI text-completion technologies, which can produce creditable content cheaply and quickly. We should note that this scenario, where an emerging technology threatens to undermine assessments, is not without some comparable historical counterparts. It was initially imagined that the advent of the internet posed an existential threat to certain kinds of assessments. However, what has emerged since then is a total revolution in the way that we teach, stemming from careful, evidence-based consideration of how we can use this platform to enhance the quality of our teaching and to improve the student experience. It would be hard to imagine delivering a modern course without using the internet in some way. On the basis of startling and exciting recent advances in the capabilities of AI technologies, notwithstanding the results of the present work, we would argue that AI is well-placed to instigate dramatic and profound changes to the way that we teach and assess students, forming an indispensable component of a new ethos within which we design and deliver teaching. 

Within the coming months, this area of research will likely undergo an explosion. Indeed, at the time of writing, we are on the cusp of the release of GPT-4. It is the view of the authors that within the next twelve months, the more pioneering students or \emph{'Mavens'} \cite{gladwell2003tipping} will realise the potential of these tools, and this news will steadily spread around student cohorts. Within two years, most higher education practitioners will be aware of how powerful these tools are and may have unwittingly marked work created in part by AI in a co-pilot manner. The potential for this technology to produce extremely accurate, high-quality content to almost any brief imaginable will surely lead to a new stage in the development of education. In the meantime, we must prepare for the AI revolution that is already underway, to ensure the fidelity of current forms of assessment.
\end{section}
\printbibliography

@inproceedings{gtp3-paper,
 author = {Brown, Tom and Mann, Benjamin and Ryder, Nick and Subbiah, Melanie and Kaplan, Jared D and Dhariwal, Prafulla and Neelakantan, Arvind and Shyam, Pranav and Sastry, Girish and Askell, Amanda and Agarwal, Sandhini and Herbert-Voss, Ariel and Krueger, Gretchen and Henighan, Tom and Child, Rewon and Ramesh, Aditya and Ziegler, Daniel and Wu, Jeffrey and Winter, Clemens and Hesse, Chris and Chen, Mark and Sigler, Eric and Litwin, Mateusz and Gray, Scott and Chess, Benjamin and Clark, Jack and Berner, Christopher and McCandlish, Sam and Radford, Alec and Sutskever, Ilya and Amodei, Dario},
 booktitle = {Advances in Neural Information Processing Systems},
 editor = {H. Larochelle and M. Ranzato and R. Hadsell and M.F. Balcan and H. Lin},
 pages = {1877--1901},
 publisher = {Curran Associates, Inc.},
 title = {Language Models are Few-Shot Learners},
 url = {https://proceedings.neurips.cc/paper/2020/file/1457c0d6bfcb4967418bfb8ac142f64a-Paper.pdf},
 volume = {33},
 year = {2020}
}

@inproceedings{gpt-3-indistinguish,
    title = "All That{'}s {`}Human{'} Is Not Gold: Evaluating Human Evaluation of Generated Text",
    author = "Clark, Elizabeth  and
      August, Tal  and
      Serrano, Sofia  and
      Haduong, Nikita  and
      Gururangan, Suchin  and
      Smith, Noah A.",
    booktitle = "Proceedings of the 59th Annual Meeting of the Association for Computational Linguistics and the 11th International Joint Conference on Natural Language Processing (Volume 1: Long Papers)",
    month = aug,
    year = "2021",
    address = "Online",
    publisher = "Association for Computational Linguistics",
    url = "https://aclanthology.org/2021.acl-long.565",
    doi = "10.18653/v1/2021.acl-long.565",
    pages = "7282--7296",
}

@article{NYT-gpt,
 author  = {Metz, Cade},
 date    = {2020-11-24},
 title   = {Meet GPT-3. It Has Learned to Code (and Blog and Argue).},
 journal = {The New York Times},
 url     = {https://www.nytimes.com/2020/11/24/science/artificial-intelligence-ai-gpt3.html}
}

@inproceedings{detect-ai,
 author = {Pillutla, Krishna and Swayamdipta, Swabha and Zellers, Rowan and Thickstun, John and Welleck, Sean and Choi, Yejin and Harchaoui, Zaid},
 booktitle = {Advances in Neural Information Processing Systems},
 editor = {M. Ranzato and A. Beygelzimer and Y. Dauphin and P.S. Liang and J. Wortman Vaughan},
 pages = {4816--4828},
 publisher = {Curran Associates, Inc.},
 title = {MAUVE: Measuring the Gap Between Neural Text and Human Text using Divergence Frontiers},
 url = {https://proceedings.neurips.cc/paper/2021/file/260c2432a0eecc28ce03c10dadc078a4-Paper.pdf},
 volume = {34},
 year = {2021}
}

@inproceedings{SA-medical-qa,
    title = "Towards Automating Healthcare Question Answering in a Noisy Multilingual Low-Resource Setting",
    author = "Daniel, Jeanne E.  and
      Brink, Willie  and
      Eloff, Ryan  and
      Copley, Charles",
    booktitle = "Proceedings of the 57th Annual Meeting of the Association for Computational Linguistics",
    month = jul,
    year = "2019",
    address = "Florence, Italy",
    publisher = "Association for Computational Linguistics",
    url = "https://aclanthology.org/P19-1090",
    doi = "10.18653/v1/P19-1090",
    pages = "948--953"
}

@article{legalAdvice,
  title={Improving access to justice with legal chatbots},
  author={Queudot, Marc and Charton, {\'E}ric and Meurs, Marie-Jean},
  journal={Stats},
  volume={3},
  number={3},
  pages={356--375},
  year={2020},
  publisher={MDPI}
}

@article{student-feedback,
  title={Generative Grading: Near Human-level Accuracy for Automated Feedback on Richly Structured Problems},
  author={Malik, Ali and Wu, Mike and Vasavada, Vrinda and Song, Jinpeng and Coots, Madison and Mitchell, John and Goodman, Noah and Piech, Chris},
  journal={arXiv preprint arXiv:1905.09916},
  year={2019}
}

@book{gladwell2003tipping,
  title={The Tipping Point: How Little Things Can Make a Big Difference},
  author={Gladwell, M.},
  isbn={9781587243936},
  lccn={2002193375},
  series={Compass Series},
  url={https://books.google.co.uk/books?id=w6V\_PwAACAAJ},
  year={2003},
  publisher={Wheeler Pub.}
}

@article{essayAutoMarkSherman,
title = {State-of-the-art automated essay scoring: Competition, results, and future directions from a United States demonstration},
journal = {Assessing Writing},
volume = {20},
pages = {53-76},
year = {2014},
issn = {1075-2935},
doi = {https://doi.org/10.1016/j.asw.2013.04.001},
url = {https://www.sciencedirect.com/science/article/pii/S1075293513000196},
author = {Mark D. Shermis}
}

@article{autoEssayLitReview,
  title={An automated essay scoring systems: a systematic literature review},
  author={Ramesh, Dadi and Sanampudi, Suresh Kumar},
  journal={Artificial Intelligence Review},
  pages={1--33},
  year={2021},
  publisher={Springer}
}

@article{VLEimpact,
  title={Why hasn’t technology disrupted academics’ teaching practices? Understanding resistance to change through the lens of activity theory},
  author={Blin, Fran{\c{c}}oise and Munro, Morag},
  journal={Computers \& Education},
  volume={50},
  number={2},
  pages={475--490},
  year={2008},
  publisher={Elsevier}
}

@article{aiCoPilotEssays,
  title={How to cheat on your final paper: Assigning AI for student writing},
  author={Fyfe, Paul},
  journal={AI \& SOCIETY},
  pages={1--11},
  year={2022},
  publisher={Springer}
}

@article{gpt3AcademicPaper,
  title={Automated Essay Writing: An AIED Opinion},
  author={Sharples, Mike},
  journal={International Journal of Artificial Intelligence in Education},
  volume={32},
  number={4},
  pages={1119--1126},
  year={2022},
  publisher={Springer}
}

@article{MOOCcompletion,
  title={Massive open online course completion rates revisited: Assessment, length and attrition},
  author={Jordan, Katy},
  journal={International Review of Research in Open and Distributed Learning},
  volume={16},
  number={3},
  pages={341--358},
  year={2015},
  publisher={Athabasca University Press (AU Press)}
}

@misc{hall_2021,
 title={Undergraduate module averages 2020 - A Freedom of Information request to University of Durham},
 url={https://www.whatdotheyknow.com/request/undergraduate_module_averages_20_25},
 journal={WhatDoTheyKnow},
 author={Hall, T}, 
 year={2021}, 
 month={07}
 }
\end{document}